\documentclass[twocolumn,showpacs,preprintnumbers,amsmath,amssymb,prl]{revtex4}

\usepackage{graphicx}% Include figure files
\usepackage{dcolumn}% Align table columns on decimal point
\usepackage{bm}% bold math

\begin{document}
\input{epsf}

\title{Ultra-High Energy Quenching of the LPM Effect: Implications for
GZK-Violating Events}

\author{John P. Ralston$^1$, Soebur Razzaque$^2$ and Pankaj Jain$^3$}

\affiliation{$^1$Department of Physics and Astronomy, University of
Kansas, Lawrence, Kansas 66045}

\affiliation{$^2$Department of Astronomy \& Astrophysics, Penn State
University, University Park, Pennsylvania 16802}

\affiliation{$^3$Physics Department, I.I.T. Kanpur, India 208016}

\date{\today}

\begin{abstract} 
Cosmic rays in the regime of $10^{19}-10^{20}$ eV should not exist,
according to the GZK bound.  Conflicts between data and the bound have
led to suggestions that laws of physics must be modified.  We find
that physical processes neglected in the early history terminate the
LPM effect, important in calibrating the energy of electromagnetic and
hadronic showers in the GZK energy regime.  The processes neglected
are the direct losses of electrons and positrons, overlooked by LPM
and greatly augmented by ultra-high energy circumstances.  We present
an exact formula for direct energy losses of $e^{+}e^{-}$ pairs in
electromagnetic showers.  Numerical integration in current hadronic
models shows that actual energy losses are always substantially larger
than LPM estimates, but the energy at which crossover occurs depends
on models.
\end{abstract}


\maketitle

Some thirty years ago, Greissen, Zatsepin and Kuzmin (GZK) \cite{gzk}
found that cosmic rays above about $5 \times 10^{19}$ eV should not be
observed.  Yet since its inception the GZK bound has never been
confirmed, as evidenced by observation of dozens, if not hundreds of
anomalous events.  Extraordinary contradictions are indicated by
angular correlations \cite{angcorrels} of GZK-violating showers and
cosmologically distant sources.  Experimental facilities such as HiRES
and AUGER are poised to explore the question with vastly more data.
The conflict between the data and the bound has led many theoretical
workers to suggest that laws of physics must be modified.  Suggestions
of new physics include exotic particles, magnetic monopoles
\cite{weiler}, neutrinos with large cross sections from low-scale
gravity \cite{Jain:2000pu}, and even the rejection of Lorentz
invariance \cite{lorentzvio}.

High energy air showers tend to be measured better with increasing
energy, due to increased signal-to-noise.  However the calibration of
shower energies depends on details of the shower evolution.  Hadronic
showers have some model dependence, but the electromagnetic components
are thought to be well-understood.  Many years ago, Landau and
Pomeranchuck (LP) \cite{LPM} showed that high energy electromagnetic
interactions in matter destructively interfere, effectively increasing
the interaction length.  Migdal (M) \cite{LPM} gave formulas smoothly
interpolating between different coherence regimes.  Monte Carlo codes
incorporate these effects, currently believed to extend to
asymptotically high energies.  The question of ultra-high energy
calibration involves what and how much to believe from LPM.

The LPM calculations hinge on dominance by a concept called the
``formation length'' of bremsstrahlung.  In a peculiar remark,
Galitsky and Guerevich (GG) \cite{galitsky64} noted an instance where
the absorption length might be less than the coherence length.  Then
photons would be absorbed before they were fully emitted.  GG were
discussing $e^{+}e^{-}$ pair production via the Bethe-Heitler (BH)
cross section.  Apparently GG missed the suppression of pair
absorption by LPM effects, which decreases the pair-production rate
self-consistently.  In fact the LPM regime extends to arbitrarily high
energies, {\it if treated as an electrodynamics problem in isolation.}
Yet nobody working before 1960 could anticipate the evolution of high
energy {\it hadronic } physics.

We highlight a rather subtle point: the hadronic component of
electrons, positrons and photons.  High energy electrons undergo two
independent causes of electromagnetic energy loss: the {\it radiative
losses} and the {\it direct} losses.  {\it Radiative} losses generate
amplitudes proportional to the acceleration $\dot \beta$.  {\it
Direct} losses exist as $\dot \beta \rightarrow 0$, and were not
mentioned by LPM. Direct losses traditionally start with the
equivalent photon density $dN_{\gamma}/dx_{\gamma} $ per Feynman
$x_{\gamma} $ in the uniformly moving electron's fields.  When such
photons interact with cross section $\sigma_{\rm tot}$, then the
equivalent energy loss per unit distance $z$ is given by
\begin{eqnarray} 
\frac{dE_{\rm equivalent}}{dz}= E\int
dx \, x\, \frac{dN_{\gamma}}{dx} \, n \sigma_{\rm tot} (xE).
\label{eqivs}
\end{eqnarray} 
The electron then has hadronic interactions because $\sigma_{\rm tot}$
has hadronic contributions, commonly called photonuclear reactions.
That is, the physical electron cannot be separated from a comoving
cloud of photons, and the physical comoving photons cannot be
separated from their comoving component of hadrons.

Numerous hadronic interactions are carefully implemented in current
air-shower Monte carlo codes.  For muons and tau leptons the direct
photonuclear losses are larger than the radiative losses \cite{reno},
and are included via equivalent photons in current codes: but not for
the electron.  Meanwhile the photonuclear reactions are not
LPM-suppressed because they occur on very short hadronic distances and
time scales.  Absorption of photons also affects the LPM radiative
losses.  Since electron and positron pairs, plus photons, are the
primary component of electromagnetic shower cascades, the photonuclear
losses by electrons may have an observable effect in shower evolution.

Shower energy calibration also involves detailed features of the
detector response, which is not our task.  Our task here is the
termination of the LPM regime above a certain energy.  The exact
electromagnetic energy loss comes from the Poynting vector,
\begin{eqnarray}
\frac{dE_{\rm tot}}{dz} = \frac{b_{\rm min}}{2\pi} \int d\phi d\omega
\, [{\vec E}_{\omega}(b_{\rm min}) \times {\vec B}_{\omega}^{*}(b_{\rm
min})]
\end{eqnarray}
Here ${\vec E}_{\omega}(b), \, {\vec B}_{\omega}(b) $ are the electric
and magnetic fields in Fourier $\omega$ space evaluated at transverse
spatial coordinate $\vec b$.  The trajectory of a fast charge has a
series of small angle kinks distributed along a nearly straight path.
Up to a phase, the radiation fields from a particular kink are given
by
\begin{eqnarray}
\vec RE_{\rm rad}=Q\frac{ \hat n \times \hat n \times \Delta \vec
\beta}{1- \hat n \cdot \vec \beta},  
\label{kink}
\end{eqnarray}
where $\hat n$ is a unit vector pointing to the observation point and
$\Delta \vec \beta$ is the change in velocity at the kink in units of
$c$.  Kinks separated by less than the formation length of radiation
add their fields with destructive interference.  In free space the
formation length $L_{\rm o}\sim \gamma^{2} /\omega$ for a charge with
Lorentz boost factor $\gamma$, while in the LPM regime $L\sim
\sqrt{\omega}L_{\rm o}/\gamma.$

The kink field [Eq. (\ref {kink})] transforms like a boosted dipole.
Meanwhile the ``kinematic'' monopole fields of the boosted Coulomb
charge are azimuthally symmetric about the trajectory.  By
orthogonality over the azimuthal integral, there is zero net
interference between the energy losses of the boosted Coulomb fields
and the radiation fields:
\begin{eqnarray} 
\frac{dE}{dz}=\frac{dE_{\rm radiative}}{dz}+\frac{dE_{\rm
direct}}{dz},
\end{eqnarray}

We set aside the radiative losses for now, and turn to the direct
term.  The fields of a uniformly moving charge are given by
\begin{eqnarray}
|E_{\perp\,\omega}| &=& \frac{q\omega}{v^2\epsilon_{\omega}\gamma_{\rm
m}} \sqrt{\frac{2}{\pi}} \, K_1 \left(\frac{b\omega}{v\gamma_{\rm
m}}\right) \nonumber \\ |E_{\parallel\,\omega}| &=&
\frac{-iq\omega}{v^2\epsilon_{\omega} \gamma_{\rm m}^2}
\sqrt{\frac{2}{\pi}} \, K_0 \left(\frac{b\omega} {v\gamma_{\rm
m}}\right)
\label{exact-field}
\end{eqnarray}
where $K_{0}$ and $K_{1}$ are modified Bessel functions.  The
effective Lorentz boost factor in the medium, denoted by $\gamma_{\rm
m}$, is defined as
\begin{eqnarray} 
\frac{1}{\gamma_{\rm m}} = \sqrt{1-\beta^2\epsilon_{\omega}}
\cong \sqrt{\frac{1}{\gamma^2} + 1-\epsilon_{\omega}}.
\label{gamma-medium} 
\end{eqnarray}

The corresponding formula for the energy loss is
\begin{eqnarray} 
\frac{dE}{dz} &=& \frac{2}{\pi}\,\alpha {\rm Re} \int_0^{\infty}
d\omega\,\frac{i\omega^2b_{\rm min}}{\epsilon_{\omega} \gamma_{\rm m}^*}
\left[\frac{1}{\epsilon_{\omega}}-\beta^2\right] \times \nonumber \\
&& K_1 \left(\frac{b_{\rm min}\omega}{\gamma_{\rm m}^*}\right) K_0
\left(\frac{b_{\rm min}\omega}{\gamma_{\rm m}}\right),
\label{sigmaloss} 
\end{eqnarray}
attributed to Fermi \cite{fermi40}.  Here $b_{\rm min} $ is the
minimum impact parameter for which the loss subprocess proceeds.

The Fermi formula is rather more sophisticated than the equivalent
photon approximation.  Replacement of $1/\epsilon-1 \rightarrow
(\epsilon-1) \rightarrow \sigma$ plus the limit $\omega \rightarrow
0,$ are needed to find the approximation Eq. (\ref{eqivs}).
Equivalent photons can be used to calculate various subprocesses
$d\sigma$, but only at leading-log order.  When subprocesses are
pursued beyond leading order certain re-summations are needed, which
depend on the process.  But there are no restrictions on Poynting's
theorem, even in the quantum domain.  All the terms needed for energy
loss are summed in the ``Fermi formula'' Eq. (\ref{sigmaloss}), which
is deeper than it may appear at first sight.

As a consequence, there are two distinct regimes:
\begin{eqnarray}
\frac{1}{\gamma^{2} } \gg |\epsilon_{\omega}-1|\:\:\:
\rightarrow \frac{\omega^{2}}{\gamma^{2}}< \frac{1}{b_{\rm min}^{2}}
\;; \label{limit1} \\
\frac{1}{\gamma^{2}} \ll |\epsilon_{\omega}-1|\:\:\:
\rightarrow \omega^{2}
|\epsilon_{\omega}-1| < \frac{1}{b_{\rm min}^{2}} \;.
\label{limit2}
\end{eqnarray}
We use the ``absorption length''$1/(\omega \,{\rm Im}
[\epsilon_{\omega}])=L_{\rm abs}$ as a characteristic length, of order
$1/n \sigma_{\rm tot} $.  The regime of Eq. (\ref{limit1}) corresponds
to free space, or any other cases of negligible medium effects.  The
Bessel functions then permit $\omega < m \gamma$, which is the {\it
kinematic maximum}.  Then $dE/dz \sim E$ for the familiar case of
$L_{\rm abs}$ being constant.  In the LPM model $1/L_{\rm abs}\sim
\sqrt{\omega}/E$, $\omega_{\rm max}$ is the kinematic maximum at high
enough energies, and $dE/dz \sim \sqrt{E}\;.$ In contrast, the other
regime of Eq. (\ref{limit2}) is dominated by absorption in the medium,
and $\omega_{\rm max}\sim m/\sqrt{|\epsilon_{\omega}-1|}$.  The
corresponding loss integral is constant if (say) $L_{\rm abs}$ is
constant.  The crossover between the two regimes occurs when the
absorption maximum falls below the kinematic maximum, namely
\begin{eqnarray}
\gamma^{2}> \frac{1}{ |\epsilon_{\omega}-1|} \rightarrow absorption \:
  dominates .
\label{critical}
\end{eqnarray}

Mutiplying both sides of Eq. (\ref{critical}) by wavelength $\lambda$,
the regime of absorptive dominance corresponds to $\gamma^{2}\lambda >
L_{\rm abs}$.  The left-hand side is the free-space formation length,
which ceases to apply when it exceeds the absorption length.  This is
just the criterion for absorption dominance cited by order of
magnitude formulas by GG \cite{galitsky64}.

We continue by noting that absorption and propagation are linked by
{\it unitarity} and dispersion relations relating the real and
imaginary parts of propagators.  Indeed the renormalized photon
propagator in a medium is elegantly coded into the {\it permittivity}
$\epsilon_{\omega}$ appearing in $\gamma_{\rm m}$ and Eq.
(\ref{sigmaloss}).  By definition
\begin{eqnarray} 
\frac{1}{\epsilon_{\omega}\omega^{2}-\vec k^{2}}=
\frac{1}{\omega^{2}-\vec k^{2}-\Sigma}, \,
\end{eqnarray}
where $\Sigma$ is the irreducible self-energy from summing bubble
diagrams.  (The ``pole model '' for $\epsilon_{\omega}$ of
non-relativistic tradition is one example.)  Either symbol,
$\epsilon_{\omega}$ or $\Sigma$ is equally valid to parametrize the
propagator, with the identity relating them $\epsilon_{\omega}-1=
-\frac{\Sigma}{\omega^{2}}$.  In this form the optical theorem is
\begin{eqnarray}
{\rm Im} [\epsilon_{\omega}] = \frac{n \sigma_{\rm tot}
(\omega)}{\omega},
\label{optical}
\end{eqnarray}
where $n$ is the number density of scatterers with photon total cross
section $\sigma_{\rm tot}(\omega)$.  Given $\sigma_{\rm tot}$, then
the full dielectric constant follows by dispersion relations, $${\rm
Re} [\epsilon(\omega) ]= 1+ \frac{1}{ \pi } {\cal P} \int d \omega' \,
\frac{ {\rm Im} [ \epsilon_{\omega'}] }{\omega' -\omega},$$ where
${\cal P} $ is the principal value.

We turn to the total cross section $\sigma_{\rm tot} (\omega)$.  In
the LPM model the pair production cross section is \cite{LPM}
\begin{eqnarray}
\sigma_{\rm LPM} (\omega)=\frac{7}{9}\frac{m_e^2}{n E_s \sqrt{\omega
X_{\rm o}}}
\label{lpmsigma}
\end{eqnarray}
where $E_{\rm s}=21.2$ MeV is the scattering energy scale and $X_{\rm
o}$ is the radiation length.  The corresponding dielectric properties
are
\begin{eqnarray}
\epsilon_{\omega,\,{\rm LPM}} = 1+\frac{7}{9}
\frac{m_e^2}{E_s \sqrt{X_{\rm o}}} \frac{1}{\pi\sqrt{\omega^3}}
\,{\rm ln}\left( \frac{1+\sqrt{\frac{\omega-i\varepsilon}
{E_{\rm LPM }}}}{1-\sqrt{\frac{\omega-i\varepsilon}
{E_{\rm LPM }}}} \right).
\label{lpm-dielectric}
\end{eqnarray}
where $E_{\rm LPM }$ is the LPM scale.  Note that $\sigma_{\rm LPM}
(\omega)$ decreases with $\omega$, as necessary for the
self-consistency of the LPM argument.

Meanwhile photonuclear contributions increase with $\omega$.  At some
high energy, the LPM-model absorption cross section can only be a
small fraction of $\sigma_{\rm tot}$.  This is because hadronic total
cross sections typically scale like $\sigma \sim s^{a}$, where $s=2
m_{p}E$ is the center of mass energy-squared on a target of mass
$m_{p}$.  Two explanations for this dependence are Regge/Pomeron
theory, and the observed growth of small-$x$ parton distributions
$q(x)$.  Data from HERA shows that $xq(x) \sim x^{-0.4} $ to a good
approximation, yielding $\sigma_{\rm tot}\sim E^{0.4}$.  Growth with
energy is simply due to the increased number of partons, and occurs in
just the same way for the neutrino-nucleon cross section
\cite{nucross}.  The 1998-2001 photon cross section by Donnachie and
Landshoff (DL) \cite{DL} is a highly cited example:
\begin{eqnarray} 
\sigma^{\gamma p}_{\rm DL} (E) &=& 0.00016 \left(\frac{2E}{\rm GeV} 
\right)^{0.4372} + 0.067 \left(\frac{2E}{\rm GeV} 
\right)^{0.0808} \nonumber \\ && + 0.129 \left(\frac{2E}{\rm GeV} 
\right)^{-0.4525} \,{\rm mb}.
\label{DL}
\end{eqnarray}
We will use DL as an up to date baseline for $\sigma_{\gamma}$,
although all extrapolations into the UHE regime should be used with
caution.  The corresponding dielectric properties are
\begin{eqnarray}
\epsilon_{\omega, {\rm DL}} = 1- \Sigma_j \frac{nC_j}
{\pi \omega_{\rm min}^{1-d_j}} F \left(1, 1-d_j, 2-d_j, 
\frac{\omega-i \varepsilon }{\omega_{\rm min}} \right)
\label{DL-epsilon}
\end{eqnarray}
where $\omega_{\rm min}\sim$ GeV,
$F(\alpha,\,\beta,\,\gamma,\,\delta)$ is the hypergeometric function,
$C_{j} $ are the prefactors and $d_j$ are the powers of $\omega$ in
Eq. (\ref{DL}).  Cross sections are expected to grow until parton
saturation sets in, after which asymptotic behavior such as the
Froissart bound is expected.  The DL total cross section reaches a
modest geometric value (40 mb) at $E\sim 10^{21}$ eV, and
extrapolations far above that value might be slowed to $log^{2}(E)$ or
similar asymptotic behavior.

We are not aware of a model for $\sigma_{\rm tot}(\omega)$ which does
not rise in the high energy regime.  We compared the 1981 fit of
Bezrukov and Bugaev (BB) \cite{DL} such that Froissart bound
dependence sets in at about 50 GeV. The BB fit is
\begin{eqnarray}
\sigma^{\gamma N}_{\rm BB}(E) = 114.3+1.67 \,{\rm ln}^2
\left( \frac{0.0213\,E}{\rm GeV} \right) \,\mu{\rm b}.
\label{bb-photonuclear}
\end{eqnarray}
We use the BB model to compare model-dependence.  We found
$\epsilon_{\omega}$ analytically by dispersion relations in terms of
poly-log functions.  A more useful representation as asymptotic series
is:
\begin{eqnarray}
\epsilon_{\omega, {\rm BB}} &=& 1+ \frac{114.3n}{\pi\omega}\,{\rm ln}
\left(\frac{-\omega_{\rm min}}{\omega-i\varepsilon} \right) + 
\nonumber \\ && \frac{1.67n}{3\pi\omega} \left[ 95.01 +
54.32 \,{\rm ln} \left(\frac{-1}{\omega-i\varepsilon} \right) +
\right. \nonumber \\ && \left.
11.55\,{\rm ln}^2 \left(\frac{-1}{\omega-i\varepsilon} \right) +
{\rm ln}^3 \left(\frac{-1}{\omega-i\varepsilon} \right) \right]. 
\end{eqnarray}

\begin{figure} [htb]
\vskip -0.3cm
\centerline{\epsfxsize=3.5in \epsfbox{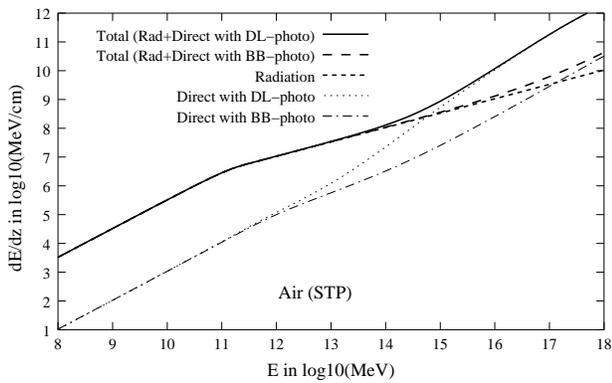}}
\vskip -0.5cm
\caption{Total energy loss ($dE/dz$) of electrons in air at sea level
($X_{\rm o} = 30420$ cm) assuming standard LPM radiative losses.
Losses with the DL model are indicated by the solid line, with the BB
model (long-dashed) shown for comparison.  Short-dashed line shows the
LPM-suppressed radiative losses, which turn over at $E_{\rm LPM} \sim
2.3 \times 10^{11}$ MeV. The direct losses of both the DL- and BB-
models terminate the LPM regime (radiative losses) above $5 \times
10^{20}$ and $10^{23}$ eV, respectively.}
\label{fig:dedzair}
\end{figure}

Potential effects for GZK are illustrated in Fig. \ref{fig:dedzair},
based on numerical integration of Eq. (\ref{sigmaloss}) for air at sea
level ($X_{\rm o}\sim 30420$ cm; $b_{\rm min}=1/m_{\pi}$).
Conventional radiative losses (also shown) reveal the LPM downturn
above $E_{\rm LPM}^{\rm Air} \sim 2.3 \times 10^{11}$ MeV.  The
DL-based calculation becomes non-negligible above about $10^{19}$ eV
and exceeds the radiative losses above $5 \times 10^{20}$ eV.  The
BB-based losses are negligible through this energy range, but also
exceed the radiative losses above $10^{23}$ eV.  Eventually any
$\sigma_{\rm tot}(\omega)$ model on the market will terminate the LPM
regime.

The onset of large direct losses of the DL-model (Fig.
\ref{fig:dedzair}) coincides alarmingly with the energy range where
the GZK bound is confronted.  As far as known, the actual cross
sections could easily be larger than DL's, causing the regime to occur
even earlier.  However we have not finished with the repercussions,
because there remain the radiative losses postponed earlier.  For any
given $x_{\gamma} =\omega/E$ fraction, the LPM coherence length
$L_{\rm LPM} \sim \sqrt{\gamma/x}$ increases indefinitely with
$\gamma$.  Meanwhile the photonuclear absorption length $L_{\rm
abs}\sim (\gamma x)^{-a} $ decreases.  When $L_{\rm LPM}>L_{\rm abs}$,
the radiative losses are no longer comoving and superposing ``on top
of one another'' near the speed of light.  Instead, the radiation from
each kink is absorbed before it reaches the next kink.  Thus quenching
of radiative LPM occurs just where the direct losses become noticable
\cite{inprep}.  This is just the same physics as the experimentally
observed \cite{klein} low energy ``polarization effect'' of
Ter-Mikaelian \cite{fermi40}, postponed to a regime of large enough
$\epsilon-1$.  We have nevertheless prepared Fig. \ref{fig:dedzair}
using the {\it standard LPM radiative losses} to facilitate
comparison.  At an order of magnitude energy (say) above the crossover
point, the direct losses dominate anyway, so that it does not matter
that the LPM formulas are an overestimate.  The crossover from LPM
radiative to direct absorptive regime is inevitable.  Unfortunately
the uncertainties of $\sigma_{\rm tot}$ make the onset energy of
absorptive dominance very uncertain.

Figure \ref{fig:dedzair} has implications for the calibration of air
shower energies.  The maximum number of shower particles $N_{\rm max}$
increases with primary energy.  The effects of LPM are to slow this
increase: for photon-initiated showers at $3 \times 10^{20}$ eV,
$N_{\rm max}\sim 1.3 \times 10^{11}$ with LPM suppression, and $N_{\rm
max}\sim 2.2\times 10^{11}$ with LPM turned off \cite{cillis99a}.
Adding direct losses is akin to turning LPM off (Fig.
\ref{fig:dedzair}).  However observational constraints on shower
energies include not only $N_{\rm max}$, but also the {\it shape} of
the shower based on Greissen-type shower profiles.  There is no direct
way to go from a particular cross section to the entire shower
profile, which incorporates every possible particle reaction.  Indeed,
the direct losses we study apply not only to the electron, but to
every single charged particle!

We conclude that extrapolations of energy calibrations to $10^{19}$ eV
and above are incomplete and model-dependent.  Large photonuclear
cross sections violating no existing physical principles may have
drastic effects.  The effects on observations remain to be determined,
and likely cannot be resolved without a concensus among observers.
With hundreds of papers citing a looming GZK crisis, we find a new
burden of proof must be borne by those citing violation of the GZK
bound.  Observations in the regime of $10^{20}$ eV will likely be
controversial for some time to come.

Work supported in part under Department of Energy grant.  We thank
Graham Wilson and Doug McKay for comments. SR was partially supported
by NSF AST0098416 grant.

\end{document}